\documentclass[12pt]{iopart}

\usepackage{graphicx}
\usepackage{pgfplots}
\usepackage{amssymb}
\usepackage{tikz}
\usepackage{graphicx}
\usepackage{pgfplots}
\usetikzlibrary{shapes,arrows}
\usepackage{textcomp}
\usepackage[utf8]{inputenc}
\usepackage[english]{babel}
\usepackage{amsthm}
\newtheorem{theorem}{Theorem}
\newtheorem{corollary}{Corollary}
\newtheorem{definition}{Definition}
\newtheorem{assumption}{Assumption}
\newtheorem{lemma}{Lemma}
\begin{document}

\tikzstyle{block} = [draw, rectangle, 
    minimum height=3em, minimum width=6em]

\tikzstyle{input} = [coordinate]
\tikzstyle{output} = [coordinate]
\tikzstyle{pinstyle} = [pin edge={to-,thin,black}]

\title{Robust Stability of Uncertain Quantum Input-Output Networks}

\author{Peyman Azodi\footnote{Current address: Department of Chemistry, Princeton University, Princeton, USA. }, Alireza Khayatian, Peyman Setoodeh}

\address{Department of Electrical and Computer Engineering, Shiraz University, Shiraz, Iran}
\ead{pazodi@princeton.edu}
\vspace{10pt}
\begin{indented}
\item[]May 2020
\end{indented}

\begin{abstract}
This paper presents a systematic method to analyze stability and robustness of uncertain Quantum Input-Output Networks (QIONs). A general form of uncertainty is introduced into quantum networks in the SLH formalism. Results of this paper are built up on the notion of \textit{uncertainty decomposition} wherein the quantum network is decomposed into nominal (certain) and uncertain sub-networks in cascade connection. Sufficient conditions for robust stability are derived using two different methods. In the first approach, a generalized small-gain theorem is presented and in the second approach, robust stability is analyzed within the framework of Lyapunov theory. In the second method, the robust stability problem is reformulated as feasibility of a Linear Matrix Inequality (LMI), which can be examined using the well-established systematic methods in the literature.
\end{abstract}

%
\vspace{2pc}

\section{Introduction}\label{intro}

\par Quantum technology has led to new ways of information processing \cite{dowling2003quantum}. Theoretical studies have demonstrated potential quantum supremacy in computation \cite{harrow2017quantum}. Large-scale quantum networks provide a general mathematical framework to model, design, and analyze various quantum information processing circuits in different implementation frameworks for future advancements. 

\par Linear QIONs are formed from localized linear quantum optical components interacting through bosonic coherent fields. Input-output theory was first proposed by Gardiner and Collett in the $1980$s as a mathematical formalism to model QIONs \cite{PhysRevA.31.3761}. Around the same time, Hudson and Parthasarthy developed a state-space formalism to model interacting linear quantum elements, known as Quantum Stochastic Differential Equations (QSDEs) \cite{hudson1984quantum}. After about a decade, the concept of quantum feedback control, which had been previously proposed in \cite{Warren1581}, was brought into quantum networks by Wiseman and Milburn \cite{PhysRevA.49.4110}, and latter by Lloyd \cite{PhysRevA.62.022108}. Control-theoretic study of QIONs began to attract the attention of a larger community of researchers from early $2000$s, when Kimura and Yanagisawa proposed the transfer-function approach to analyze quantum networks \cite{yanagisawa2003transfer,yanagisawa2003transfer2}. Then, Gough and James developed the {SLH} formalism, which is an algebraic structure to study QIONs beyond series connection of elements  \cite{gough2009series,gough2009quantum}. In parallel, the idea of applying robust control to quantum systems was proposed by Doherty, Mabuchi, and their colleagues \cite{stockton2004robust,doherty2000robust}. Later, James, Nurdin, and Petersen developed a concrete robust control scheme for linear QIONs \cite{james2008h}.  
\par Linear QIONs in the SLH formalism have been proved to be substantially efficient in modelling quantum information-processing circuits. The spectrum of applications of QIONs includes designing superconducting qubit networks \cite{PhysRevLett.120.213602,PhysRevA.98.013801}, Silicon semiconductor \cite{sarovar2016silicon,benito2017input,d2019optimal,doi:10.1063/5.0004777}, and entanglement dynamics in superconductors \cite{Bienfait368,zhong2019violating} as well as in NV-color centers \cite{du2016general} and their evolution via coherent feedback control \cite{Yan_2017}. Recently, an input-output theoretic approach has been proposed for quantum state transfer \cite{zeuthen2020figures}. 
 QIONs constitute an excellent framework for studying Gaussian quantum information processing by the virtue that they preserve Gaussian states \cite{weedbrook2012gaussian,levitt2018power}. Since modelling QIONs in the SLH framework provides a structured and systematic method to design and analyze large-scale networks of interacting quantum elements, it has increasingly attracted  researchers' attention in recent years. The SLH formalism has also been used to optimize photo-detection in microwave circuits \cite{Sch_ndorf_2018}.
\par Uncertainty is unavoidable in quantum networks. The fact that explicitly considering robustness against uncertainties is essential in quantum control systems was emphasized at early stages of this theory \cite{Warren1581}. Implementation of quantum information processors exhibits unavoidable and undesirable dynamics, for instance, complete purification of diamond has not been achievable and nuclear spin noise is an important limiting factor in NV-center quantum information processing \cite{DOHERTY20131}. In scalable Silicon quantum dots, decoherence caused by spin-photon coupling or inter-dot tunneling can lead to many unwanted random dynamics \cite{mi2018coherent}. Unavoidable cross-talk occurs between ions in cryogenic trapped ion systems \cite{doi:10.1063/1.5088164}. Robust control techniques can mitigate the effect of undesirable dynamics and lead to more efficient quantum information processors.  
\par As mentioned earlier, robust control theory was first used for linear QIONs in \cite{james2008h, doherty2000robust}. In \cite{doi:10.1098/rsta.2011.0527}, a sufficient condition was established for robust stability of linear quantum systems with additive quadratic uncertainty in the Hamiltonian operator. Later, result of the mentioned paper was generalized to nonlinear non-quadratic \cite{6760165} and Weyl-quantized \cite{7358693} form of perturbation in the Hamiltonian operator. In \cite{petersen2012robust}, a sufficient bounded-real lemma condition was derived for robust stability of linear quantum networks with uncertain coupling operator (L). In \cite{Petersen_2017}, a Popov stability criterion was established for quantum systems consisting of an optical cavity containing a saturated Kerr medium.
\par In this paper, two methods are presented to study robust stability of linear quantum networks with uncertainties in all three parameters (S,L,H). The presented theory is built up on the notion of uncertainty decomposition \cite{azodi2019uncertainty}, in which the uncertain network is first decomposed into nominal (certain) and uncertain sub-networks, which are connected in a cascade manner. This scheme provides a general framework for robust stability analysis of uncertain QIONs. This phenomenon, which enables us to provide robust stability analysis for networks with uncertainties in all three characterizing parameters (S,L,H), is the main distinguishing novelty of this paper. Therefore, through using the uncertainty decomposition scheme, the existing robust stability analysis methods in the literature, which consider uncertainties in just the Hamiltonian operator \cite{doi:10.1098/rsta.2011.0527} or the coupling operator \cite{petersen2012robust}, can be unified under one umbrella. In other words, the uncertainty decomposition scheme provides a general framework for robust stability analysis that encompasses a larger class on uncertain quantum networks. In the first approach, a modified small-gain condition for robust stability is presented for the decomposed quantum network. The presented small-gain theorem provides an upper bound on the uncertain parameters. In the second method, an LMI-based condition is derived for robust stability. This method provides a computationally effective and structured method to analyze robust stability for quantum information-processing purposes, which is the second novelty of this paper.

\par This paper is organized as follows. In section \ref{Brief}, introductory algebraic structure of QIONs in the SLH framework is presented. In Section \ref{sec3}, the uncertainty decomposition scheme for linear QIONs is briefly discussed. Contributions of this paper are presented in section \ref{results}. In subsection \ref{results1}, the first method, the small-gain-like approach, is demonstrated, and in \ref{Lyapunov}, the second method, the Lyapunov-based approach is explained in details.

\section{Brief Review of the \textbf{SLH} Formalism}\label{Brief}
Quantum Input-Output Networks are formed from simple linear elements, which are coupled to each other through coherent fields. In the {SLH} formalism, each element of the network is modeled using the triplet $(S,L,H)$, encoding all static and dynamic properties of the element. Based on how these elements are connected to each other, parameters of the network can be obtained from parameters of the constituents, using algebraic relations \cite{gough2009series,gough2009quantum}.
In the rest of this section, the $(S,L,H)$ parameters  and the state-space realization of linear QIONs are presented. More details on these topics can be found in \cite{combes2017slh,azodi2019uncertainty}.
\par Analogy with Open Quantum Harmonic Oscillators (OQHOs) can be used to model linear components in the {SLH} framework. State of a linear component with $n$ different frequency modes includes annihilation $a_i(t)$ and creation $a_i^{\dag}(t)$ operators for $i=1,\cdots ,n$,  defined on the underlying Fock space in the following \textit{doubled-up} form: 
\begin{equation}\label{state}
\mathop X\limits^ \cup   \doteq \left( {\begin{array}{*{20}{c}}
  X \\ 
  {{X^\# }} 
\end{array}} \right); \hspace{0.7cm}  X=(a_1(t),\cdots, a_n(t))^T, X^\#=(a_1^\dag (t),\cdots, a_n^\dag (t))^T.
\end{equation}
Based on this definition, any linear combination ($Y$) of system states in the form of
\begin{equation}
    Y = {E^ - }X + {E^ + }{X^\# }; \;\; {E^ - }\& {E^ + } \in {\mathbb{C}^{n' \times n}},
\end{equation}
can be written as:
$$\mathop Y\limits^ \cup   = \Delta ({E^ - },{E^ + })\mathop X\limits^ \cup,  $$
using the following definition for doubled-up matrix of coefficients:
\begin{equation}
\Delta ({E^ - },{E^ + }) \doteq \left( {\begin{array}{*{20}{c}}
  {{E^ - }}&{{E^ + }} \\ 
  {{E^ + }^\# }&{{E^ - }^\# } 
\end{array}} \right).
\end{equation}

An $m-$input, $m-$output linear element in the network is coupled to $m$ independent bosonic annihilation ${\rm A}_i^{in}(t)$, and creation fields ${\rm A}_i^{in\dag}(t); \; i:1,...,m$, defined on Fock spaces ${F_i}$. The corresponding output fields are denoted by ${\rm A}_i^{out}(t)$ and ${\rm A}_i^{out\dag}(t); \; i:1,...,m$. Doubled-up input and output field vectors are:
\begin{eqnarray}\label{inout}
 {{\mathop{\rm A}\limits^ \cup}^{{in}}}(t) = {\left( {{\rm A}_1^{in}(t),\cdots,{\rm A}_m^{in}(t),{\rm A}_1^{in\dag}(t),\cdots,{\rm A}_m^{in\dag}(t)} \right)^{\rm T}},\\ {{\mathop{\rm A}\limits^ \cup}^{{out}}}(t) = {\left( {{\rm A}_1^{out}(t),\cdots,{\rm A}_m^{out}(t),{\rm A}_1^{out\dag}(t),\cdots,{\rm A}_m^{out\dag}(t)} \right)^{\rm T}}.
\end{eqnarray}
For both input and output fields, if $d \rm A_i(t)$ denotes the forward-time It\"o increment of $\rm A_i(t)$, then, the following It\"o products are relevant \cite{parthasarathy2012introduction}:
\begin{eqnarray}
  d{\rm A}_i^{}(t)d{{\rm A}^\dag}_j^{}(t') = {\delta _{ij}}\delta (t-t')dt, \\ 
  d{{\rm A}^\dag}_i^{}(t)d{{\rm A}^\dag}_j^{}(t') = 
  d{\rm A}_i^{}(t)d{\rm A}_j^{}(t') =
  d{{\rm A}^\dag}_i^{}(t)d{\rm A}_j^{}(t') = 0.  
\end{eqnarray}
\par In the $SLH$ formalism, $H$ corresponds to the internal energy dynamics and the pair $(S,L)$ specifies the interface of the system to external fields:
\begin{itemize}
\item{The scattering matrix, $S \in {\mathbb{C}^{m \times m}}$, determines the input-output fields static relation and is a unitary matrix (${S^\dag }S = S{S^\dag } = I$ ).}
\item{The coupling operator, $L = C\mathop X\limits^ \cup   = \left( {\begin{array}{*{20}{c}}
  {{C^ - }},&{{C^ + }} 
\end{array}} \right)\mathop X\limits^ \cup; \; C \in {\mathbb{C}^{m \times 2n}}$,  describes how input and output fields interact with internal dynamics.}
\item{The Hamiltonian, $H$, is Hermitian and characterizes internal evolution of system's state in the absence of input fields.}
\end{itemize}
\par The following general form for the Hamiltonian operator can be used for both passive (photon number preserving) and active components (vacuum field energy is neglected and for simplicity we put $\hbar=1$):
\begin{equation} \label{a}
H = \sum\limits_{i,j = 1}^n {\left( {\omega _{ij}^{-} a_i^{\dag}{a_j} + \frac{1}{2}\omega _{ij}^ + a_i^{\dag}a_j^{\dag} + \frac{1}{2}\omega {{_{ij}^ + }^*}a_j^{}a_i^{}} \right)},
\end{equation}
with $\omega _{ij}^ - , \omega _{ij}^ +  \in \mathbb{C}$, $\omega _{ij}^-=\omega _{ji}^{-*}$ .
\\ In correspondence with this general quadratic form (\ref{a}), let us define:
\begin{eqnarray}
{\Omega _ - } = (\omega _{ij}^ - ),{\Omega _ + } = (\omega _{ij}^ + ) \in {\mathbb{C}^{n \times n}},\label{cor0}\\
- i\mathop \Omega \limits^ \sim   =  - \Delta (i{\Omega _ - },i{\Omega _ + })\label{cor}.
\end{eqnarray}
The defined correspondence in (\ref{cor0})-($\ref{cor}$) is denoted by $H \sim \tilde \Omega$.

\par Linear QIONs can also be described by the state-space realization, $G \equiv (\tilde A,\tilde B,\tilde C,\tilde D)$. 
Consider quantum network $G \equiv (S,L,H)$, in the SLH formalism with state vector $\mathop X\limits^ \cup (t)$ defined in (\ref{state}), which is coupled to input and output field vectors ${A^{in}}(t)$  and ${A^{out}}(t)$, defined in (\ref{inout}). The following quantum stochastic differential equation describes $G$ in the state-space realization \cite{hudson1984quantum,gough2010squeezing}:
\begin{eqnarray}
  d\mathop X\limits^ \cup  (t) = \tilde A\mathop X\limits^ \cup (t) dt + \tilde Bd{\mathop{\rm A}\limits^ \cup}^{{in}}(t), \hfill \\
  d{\mathop{\rm A}\limits^ \cup}^{{out}}(t) = \tilde C\mathop {\mathop X\limits^ \cup  }\limits^{} (t)dt + \tilde Dd{ {\mathop{\rm A}\limits^ \cup}^{{in}}}(t) \hfill, 
\end{eqnarray}
where
\begin{equation}\label{e33}
  \tilde A =  - \frac{1}{2}{{\tilde C}^{\flat} }\tilde C - i\tilde \Omega ,\hspace{3mm} \tilde B =  - {{\tilde C}^{\flat} }\tilde D,\hspace{3mm} \tilde C = \Delta ({C^ - },{C^ + }),\hspace{3mm} \tilde D = \Delta (S,0), 
\end{equation}
with  $H \sim \tilde \Omega$, ${\tilde C^\flat } \doteq {J_n}{\tilde C^\dag }{J_m}$, and ${J_n} \doteq \left( {\begin{array}{*{20}{c}}
  {{I_n}}&0 \\ 
  0&{ - {I_n}} 
\end{array}} \right)$ ($I_n$ is the $n\times n$ identity matrix).
\par In this section, the {SLH} formalism and its underlying algebraic structure, the Bogoliubov Lie algebra, was briefly reviewed. This structure will be used in the proposed robust analysis scheme in the following sections.
\section{Uncertainty Decomposition in {SLH} Formalism}\label{sec3}
\par As discussed in section \ref{intro}, the {SLH} formalism has recently been successfully used for quantum information-processing purposes. This formalism provides a systematic algebraic structure to design \cite{bienfait2019phonon} and analyze \cite{pichler2016photonic} quantum circuits. In section \ref{intro}, we also mentioned some of the main sources of uncertainty in quantum networks in different quantum information-processing frameworks. It is not far from reality to conclude that uncertainty in quantum networks is inevitable. Moreover, uncertainty affects system's performance and stability, especially, in large-scale quantum networks, where small uncertainties in constituents can result in a significant deviation in the global system's performance. 
\par One pathway to apply robust synthesis and analysis techniques to large-scale quantum networks is to decompose all uncertainties in the system into an additional uncertainty block, which interacts with the certain nominal quantum network. This method is called \textit{uncertainty decomposition}. In \cite{azodi2019uncertainty}, an uncertainty decomposition scheme was proposed for quantum networks. In this scheme, uncertainties are introduced in the triplet $(S,L,H)$ in additive and multiplicative manners. Then, an uncertain quantum network can be decomposed into uncertain and nominal sub-networks in cascaded connection, denoted by $G \equiv {G_n} \triangleleft \Delta $ (Figure \ref{fig1}).

\begin{figure}[t]
\hspace{0.5cm}
\begin{tikzpicture}[auto, node distance=2cm,>=latex']
    \node [input, name=input] {};
    \node [block, right of=input,node distance=2cm] (controller) {$G$};
    \node [input, name=output,right of=controller,node distance=2cm] {};
    \draw [->] (input) -- node[name=u] {} (controller);
    \draw [->] (controller) -- node[name=cont] {} (output);
\end{tikzpicture}
$ \rightarrow{\textrm{Decomposition}} $
\begin{tikzpicture}[auto, node distance=2cm,>=latex']
    \node [input, name=input] {};
    \node [block, right of=input,node distance=2cm] (controller) {$\Delta$};
    \node [block, right of=controller,node distance=3.2cm] (system) {$G_n$};
    \node [output, right of=system, node distance=2cm] (output) {};
    \draw [->] (input) -- node[name=u] {} (controller);
    \draw [->] (controller) -- node[name=u] {} (system);
    \draw [->] (system) -- node[name=u] {} (output);
\end{tikzpicture}
\caption{Uncertainty decomposition in an uncertain linear quantum input-output network. The uncertain network $G$ is decomposed into nominal ($G_n$) and uncertainty ($\Delta$) sub-networks in cascade connection ($G \equiv {G_n} \triangleleft \Delta $). \label{fig1}}
\end{figure}

Let us consider the uncertain quantum network $G \equiv (S,L,H)$. In the proposed uncertainty decomposition algorithm, while the scattering matrix, $S$, is perturbed post-multiplicatively, $S = {S_n}\Delta S$, the coupling matrix, $L$, and the Hamiltonian, $H$, are perturbed additively, $L = {L_n} + \Delta L$ and $H = {H_n} + \Delta H$, where ${S_n}$, ${L_n}$, and ${H_n}$ denote the nominal matrices, and $\Delta S$, $\Delta L$, and $\Delta H$ are perturbations. The uncertainty sets $\overline {\Delta S} $ , $\overline {\Delta L} $  and $\overline {\Delta H}$ can be found based on the prior knowledge about the system and sources of uncertainty, which include $\Delta S$, $\Delta L$ and $\Delta H$, respectively, and characterize their properties. For instance, $\overline {\Delta L}$  might be defined as $\overline {\Delta L}  = \left\{ {\left( {{\delta _c},0} \right)\mathop X\limits^ \cup \left| {{\delta _c} \in {\mathbb{C}^{m \times n}},{{\left| {{\delta _c}} \right|}_\infty } \leqslant 1} \right.} \right\}$.
 Uncertainty sets can be different for any two distinct uncertain linear quantum systems due to their differences in elements, connections (wiring), and sources of uncertainty. Despite these differences, they must admit the following general forms:

\begin{enumerate}

\item$\forall \Delta S \in \overline {\Delta S} ;$$\Delta S$ is unitary. \label{itemi}
\item $\overline {\Delta L}  \subset \left\{ {\left( {\delta _c^ - ,\delta _c^ + } \right)\mathop X\limits^ \cup  \left| {\delta _c^ - \& \delta _c^ +  \in {\mathbb{C}^{m \times n}}} \right.} \right\}$.
\item $\forall \Delta H \in \overline {\Delta H} ;$$\Delta H$ has the mathematical form in (\ref{a})\label{itemiii}.\end{enumerate}

Nominal parameters, ${S_n}$, ${L_n}$, and ${H_n}$, along with uncertainty sets, $\overline {\Delta S} $,  $\overline {\Delta L} $, and $\overline {\Delta H}$, characterize the set of admissible systems $\overline G $:
\begin{equation}\label{adm}
\hspace{-1cm}\overline G  = \left\{ 
  \left( {{S_n}\Delta S,{L_n} + \Delta L,{H_n} + \Delta H} \right)\left| {\Delta S \in \overline {\Delta S} } \right., \hfill 
  \Delta L \in \overline {\Delta L} ,\Delta H \in \overline {\Delta H}  \hfill  \right\}.
\end{equation}
\par Based on the presented structure, the following theorem from \cite{azodi2019uncertainty} provides uncertainty decomposition algorithms in both the $SLH$ formalism and the state-space realization of quantum networks. 

\begin{theorem}[\cite{azodi2019uncertainty}]\label{th1}
Every uncertain linear quantum input-output network $G \in \overline G $, with the uncertainty structure described in (\ref{itemi})-(\ref{itemiii}) can be decomposed into two linear quantum sub-networks $\Delta $ and ${G_n}$ connected in a cascade manner, $G \equiv {G_n} \triangleleft \Delta $, as shown in Figure \ref{fig1}. Then,
\begin{enumerate}
    \item {The underlying sub-networks possess the following $(S,L,H)$ parameters:
\begin{eqnarray}\label{222}
  G \equiv ({S_n}\Delta S,{L_n} + \Delta L,{H_n} + \Delta H), \\ 
  {G_n} \equiv ({S_n},{L_n},{H_n}), \\ 
  \Delta  \equiv (\Delta S,{S_n}^\dag \Delta L,\Delta H - \mathrm{Im} (L_n^\dag \Delta L)).  
\end{eqnarray}}
\item {If the state-space realization of networks are denoted by: \begin{eqnarray*}\label{e1}
  G \equiv (\tilde A,\tilde B,\tilde C,\tilde D), 
  {G_n} \equiv ({{\tilde A}_n},{{\tilde B}_n},{{\tilde C}_n},{{\tilde D}_n}),
  \Delta  \equiv ({{\tilde A}_\delta },{{\tilde B}_\delta },{{\tilde C}_\delta },{{\tilde D}_\delta }),
\end{eqnarray*}
then, the following relation holds between state matrices ${\tilde A}$, ${\tilde A}_n$, and ${\tilde A}_\delta$:
\begin{equation}\label{a234}
\tilde A = {\tilde A_n} + ({\tilde A_\delta }  - {\tilde C_n}^\flat S_n^\dag {\tilde C_\delta }) = {\tilde A_n} + \Delta \tilde A,
\end{equation}
 More explicitly, the additive perturbation $\Delta \tilde A$ is obtained as:
\begin{equation}\label{delta}
\Delta \tilde A =  - \frac{1}{2}{\tilde C_ \delta}^\flat {\tilde C_ \delta} - i{\tilde \Omega _{\Delta H}} - {\mathrm{Re} _\beta }\{{\tilde C_n}^\flat S_n^\dag {\tilde C_ \delta}\},    
\end{equation}
where ${\tilde \Omega _{\Delta H}} \sim \Delta H$ and $\mathrm{Re} _\beta \{X\}\doteq \frac{1}{2}(X+X^\beta)$.\qed
}
\end{enumerate} 
\end{theorem}
\par Next section provides the robust stability analysis for networks in $\tilde G$ based on the uncertainty decomposition scheme of Theorem \ref{th1}. The following assumption on convexity of uncertainty sets will be used in our stability analysis.
\begin{assumption}\label{ass1}
Uncertainty sets $\overline{\Delta L}$ and $\overline{\Delta H}$ are convex, compact, and include the unique zero element of the underlying vector spaces. Consequently, if $\Delta L \in \overline{\Delta L}$, then, $\kappa\Delta L \in \overline{\Delta L}$ for all $\kappa \in [0,1]$. The same is true for $\overline{\Delta H}$.
\end{assumption}
Based on this assumption, uncertainty sets $\overline{\Delta L}$ and $\overline{\Delta H}$ are connected. This property is important to our analysis in two ways. First, connectedness of uncertainty sets is intuitive with regard to physics of uncertain linear quantum networks. Including the zero element of the underlying vector space is in accordance with the fact that when uncertainty is absent, the quantum network will operate with its \textit{nominal} performance. Moreover, connectedness indicates that uncertain values do not dis-continuously vary in uncertainty sets, but the uncertain quantum network can have all continuous realizations in $\overline{G}$. Second, this assumption will help us to build-up a concrete mathematical structure for the proposed robust analysis in the next section.

\section{Robust Stability of Uncertain Quantum Networks using Uncertainty Decomposition}\label{results}
\par In the previous section, uncertain quantum linear networks were introduced in the SLH formalism, and an uncertainty decomposition scheme was introduced. The admissible system set, $\overline G$, includes all possible realizations of the system in a domain characterized by uncertainty sets. In this section, the main concern is robust stability in the presence of uncertainties: Is there a family of triplets $(\Delta S, \Delta L, \Delta H)$ in uncertainty sets that makes a subset of quantum systems in $\overline G$ unstable? Two disjoint approaches are presented to answer this question. These approaches to robust stability analysis are applicable to different settings, which will be discussed in detail.
\par Before proceeding to the main results on robust stability analysis, robustly stable admissible quantum networks are defined. According to \cite{james2008h}, a linear quantum system $G \equiv (\tilde A,\tilde B,\tilde C,\tilde D)$ is stable, if $\tilde A$ is a Hurwitz or stable matrix (all eigenvalues of $\tilde A$ have strictly negative real parts). Therefore, robustly stable admissible set of quantum networks is defined as follows.
\begin{definition}
The admissible set of a quantum network $\overline G$ is said to be \textit{robustly stable}, if for all $G \in \overline G$, the state matrix $\tilde A$ is stable.
\end{definition}
In other words, the robust stability criterion is met when no triplet $(\Delta S, \Delta L, \Delta H)$ can destabilize $G_n$ in cascade connection \footnote{Based on Assumption \ref{ass1}, uncertainty sets include the zero element of corresponding vector spaces, thus, $G_n\in \overline{G}$. In other words, it necessitates that $G_n$ be stable.}.
\subsection{The Small-Gain-Like Approach}\label{Small}\label{results1}
In this section, a sufficient condition for robustly stable uncertain QIONs is obtained using a small-gain-like approach. The small-gain theorem is a well-known stability analysis approach for both linear and non-linear systems in control theory \cite{sastry2013nonlinear,zhou1998essentials}. Small-gain theorem provides a stability criterion for a system consisting of two interacting sub-systems in feedback connection. In the case of uncertain linear quantum systems, as shown in Figure \ref{fig1}, interacting systems are in cascade connection. Therefore our results are a generalization of the small-gain theorem for uncertain linear quantum systems in cascade connection.
\par Let us denote the set of all possible additive perturbations $\Delta \tilde A$ in (\ref{delta}) by $\overline{\Delta A}$. The following Lemma can be stated and will be used to prove the main results of this subsection:
\begin{lemma}\label{L1}
If there exists $\Delta \tilde A_1 \in \overline{\Delta A}$ such that $\tilde A _n+\Delta \tilde A_1$ is unstable. Then, there exists $\Delta \tilde A_2 \in \overline{\Delta A}$ and an integer $s$ ($0\leq s\leq 2n$), such that:
\begin{equation}
\mathrm{Re}\big\{\lambda_s (\tilde A _n+\Delta \tilde A_2)\big\}=0,
\end{equation}
where $ \lambda_i (X)$ denotes the $i$th eigenvalue of $X$.
\end{lemma}
\begin{proof}
Since $\Delta \tilde A_1$ is an element of $\overline{\Delta A}$, there exist $\Delta L_1= \tilde C _\delta  \mathop X\limits^ \cup$ in  $\overline{\Delta L}$ and $ \Delta H_1 \sim -i \tilde \Omega _{\Delta H} $ in $\overline{\Delta H}$, which are related to $\Delta \tilde A_1$ via equation (\ref{delta}). Moreover, using Assumption \ref{ass1}, for all $\kappa \in [0,1]$, we have $\kappa \Delta L_1 \in\overline{\Delta L}$ and $\kappa \Delta H_1 \in\overline{\Delta H}$. Therefore, the additive uncertainty corresponding to these uncertainty parameters,
\begin{equation} \label{ff}
    f_{\Delta \tilde A_1}(\kappa)\doteq - \frac{\kappa ^2}{2}{\tilde C_ \delta}^\flat {\tilde C_ \delta} - \kappa\big ( i{\tilde \Omega _{\Delta H}} + {\mathrm{Re} _\beta }({\tilde C_n}^\flat S_n^\dag {\tilde C_ \delta})\big ),
\end{equation}
is also a member of $\overline{\Delta A}$, for all $\kappa \in [0,1]$.
Given the hypothesis of Lemma ($\tilde A _n+\Delta \tilde A_1$ is unstable), at least one eigenvalue of $\tilde A _n+\Delta \tilde A_1$ has positive real value, thus, there exists integer $s$, such that
\begin{equation*}
    \mathrm{Re}\{\lambda_s(\tilde A _n+f_{\Delta \tilde A_1}(1))\}>0.
\end{equation*}
On the other hand, since $\tilde A_n$ is stable, we can write:
\begin{equation*}
    \mathrm{Re}\{\lambda_s(\tilde A _n+f_{\Delta \tilde A_1}(0))\}< 0.
\end{equation*}
Since (\ref{ff}) is a continuous function of $\kappa$ and eigenvalues of $f$ vary smoothly with respect to $\kappa$, the intermediate value theorem dictates that there exists $\kappa_1 \in [0,1)$ such that
\begin{equation*}
    \mathrm{Re}\{\lambda_s(\tilde A _n+f_{\Delta \tilde A_1}(\kappa_1))\}= 0.
\end{equation*}
Hence, $\Delta \tilde A_2=f_{\Delta \tilde A_1}(\kappa_1) $ is a member of $\overline{\Delta A}$ because of Assumption \ref{ass1}, which proves the Lemma.
\end{proof}
\begin{corollary}[Contraposition of Lemma \ref{L1}]\label{col2}
If equation 
\begin{equation}\label{colcon}
    \mathrm{Re}\big\{\lambda_s (\tilde A _n+X)\big\}=0,
\end{equation}
has no solution in $\overline{\Delta A}$ for all integers $0 \leq s\leq 2n$, then, $\overline{G}$ is robustly stable.\qed
\end{corollary}
\par Let us assume that additive perturbation $\Delta \tilde A$ in (\ref{delta}) admits the following norm-bounded condition:
\begin{equation}\label{nbc}
(\Delta \tilde A)^\dag \Delta \tilde A \leq \eta^2 I_{2n},
\end{equation}
for some real $\eta \geq 0$.

The following robust stability theorem can be stated.
\begin{theorem}\label{th2}
Consider the set of admissible QIONs, $\overline{G}$ in (\ref{adm}), obtained from the uncertainty decomposition algorithm in Theorem \ref{th1}. Moreover, assume that $G_n\equiv ({{\tilde A}_n},{{\tilde B}_n},{{\tilde C}_n},{{\tilde D}_n})$ is stable and $\Delta \tilde A$  satisfies the norm-bounded condition in (\ref{nbc}). Then, $\overline{G}$ is robustly stable if the following condition holds:
\begin{equation}\label{con1}
    \eta< \inf_{\omega \in \mathbb{R}}\{\underline {\sigma}(i\omega I_{2n}-\tilde A_n) \},
\end{equation}
where $\underline{\sigma}(X)$ is the smallest singular value of $X$.
\end{theorem}
\begin{proof}
Using the basic property of singular values ($|\underline{\sigma}(X+\Delta)-\underline{\sigma}(X)|\leq \overline{\sigma}(\Delta)$, where $\underline{\sigma}(.)$ and $\overline{\sigma}(.)$ denote the smallest and  the largest singular values, respectively), we can write:
\begin{equation*}
    |\underline{\sigma}(i\omega I-(\tilde A_n+\Delta \tilde A))-\underline{\sigma}(i\omega I-\tilde A_n)|\leq \overline{\sigma}(\Delta \tilde A),
\end{equation*}
therefore,
\begin{equation*}
    \underline{\sigma}(i\omega I-(\tilde A_n+\Delta \tilde A)) \geq \underline{\sigma}(i\omega I-\tilde A_n) - \overline{\sigma}(\Delta \tilde A)
     \geq \underline{\sigma}(i\omega I-\tilde A_n) - \eta,
\end{equation*}
where (\ref{nbc}) was used in the last step. If hypothesis (\ref{con1}) holds, then, 
\begin{equation*}
    \inf_{\omega \in \mathbb{R}}\{\underline{\sigma}(i\omega I-(\tilde A_n+\Delta \tilde A))\}> 0,
\end{equation*}
which implies that none of the eigenvalues of $\tilde A_n+\Delta \tilde A$ is pure imaginary (has zero real part). Equivalently, equation (\ref{colcon}) has no solution in $\overline{\Delta A}$, if (\ref{con1}) holds. Using Corollary \ref{col2}, implies that $\overline{G}$ is robustly stable.
\end{proof}
\subsection{The Lyapunov-Based Approach}\label{Lyapunov}
In this subsection, another sufficient robustly stable condition for the admissible system set $\overline{G}$ is developed, using the well-known Lyapunov stability criterion for linear systems \cite{sastry2013nonlinear} and theory of Linear Matrix Inequalities \cite{boyd1994linear}. Here, the robust stability criterion for uncertain QIONs is reformulated as a standard LMI problem, which can be solved using numerical LMI solvers. 
\par The following Schur complement Lemma will be repeatedly used in the proof of the main theorem in this subsection.
\begin{lemma}[Schur Complement \cite{zhou1998essentials}] \label{L2}
For an arbitrary square matrix, which is partitioned into diagonal-block-square sub-matrices as:
\begin{equation}
X= \left( \matrix{ X_{11} & X_{12} \cr
X_{21} & X_{22}} \right),
\end{equation}
following statements are equivalent:
\begin{enumerate}
    \item $X<0$ ($X$ is negative-definite),
    \item $X_{22}<0$ and $X_{11}-X_{12}X_{22}^{-1}X_{21}<0$.\qed
\end{enumerate}
\end{lemma}
The following theorem converts the robust stability criterion into a solvable LMI.
\begin{theorem}\label{th3}
Consider the set of admissible QIONs, $\overline{G}$ in (\ref{adm}), obtained from the uncertainty decomposition algorithm in Theorem \ref{th1}. Moreover, assume that $G_n\equiv ({{\tilde A}_n},{{\tilde B}_n},{{\tilde C}_n},{{\tilde D}_n})$ is stable and $\Delta \tilde A$  satisfies the norm-bounded condition in (\ref{nbc}). Then, $\overline{G}$ is robustly stable if either of the following conditions holds:
\begin{enumerate}
    \item \label{ii} There exists self-adjoint, positive-definite matrix $P$, such that:
    \begin{equation}
        \left( \matrix{ \tilde A_n ^\dag P+P \tilde A_n +\eta ^2 I & P \cr
P & -I} \right)< 0.
    \end{equation}
    \item \label{iiii} The following linear matrix inequality is feasible and $\zeta< \frac{1}{\eta^2}$:
    \begin{eqnarray}\label{LMI1}
    &\hspace{-2cm}\textrm{minimize}\hspace{1cm}\zeta>0\\
    &\hspace{-2cm}\textrm{subject to}\hspace{1cm} P=P^\dag>0\textrm{ and}\hspace{1cm}  
    \left( \matrix{ \tilde A_n ^\dag P+P \tilde A_n  & P & I \cr
P & -I &0 \cr I& 0& -\zeta I} \right)< 0.\label{LMI2}
    \end{eqnarray}
\end{enumerate}
\end{theorem}
\begin{proof}
Using the Lyapunov stability theory, $X$ is stable if and only if there exists a positive-definite $P$, such that $X^\dag P+P X<0$. Therefore, $\tilde A$ in (\ref{a234}), remains stable if for all $ \Delta \tilde A \in \overline{\Delta A}$, there exists a positive-definite $P$ such that:
\begin{equation}\label{211}
  (\tilde A_n+\Delta \tilde A)^\dag P+P(\tilde A_n+\Delta \tilde A)=\Delta \tilde A^\dag P+P \Delta \tilde A+\tilde A_n^\dag P+P\tilde A_n<0.  
\end{equation}
Using the fact that $\| P-\Delta \tilde A\|\geq 0$, we have:
\begin{eqnarray*}
    &(P-\Delta \tilde A)^\dag(P-\Delta \tilde A)=PP-\tilde P \Delta \tilde A-\Delta \tilde A^\dag P+\Delta \tilde A^\dag \Delta \tilde A\geq 0\\
    &\Rightarrow  P \Delta \tilde A+\Delta \tilde A^\dag P\leq PP+ \Delta \tilde A^\dag \Delta \tilde A \\
    \textrm{using (\ref{nbc})} &\Rightarrow  P \Delta \tilde A+\Delta \tilde A^\dag P\leq PP +\eta ^2 I_{2n}.
\end{eqnarray*}
Therefore, (\ref{211}) can be rewritten as:
\begin{equation}\label{3111}
   \tilde A_n^\dag P+P\tilde A_n +PP +\eta ^2 I_{2n}<0.
\end{equation}
Using Lemma \ref{L2}, this condition is equivalent to:
\begin{equation}\label{4213}
    \left( \matrix{ \tilde A_n ^\dag P+P \tilde A_n +\eta ^2 I_{2n} & P \cr
P & -I_{2n}} \right)< 0,
\end{equation}
which proves (\ref{ii}).
Moreover, inequality (\ref{4213}) can be rewritten as:
\begin{equation}
     \left( \matrix{ \tilde A_n ^\dag P+P \tilde A_n  & P \cr
P & -I_{2n}} \right)+ \left( \matrix{ I_{2n}   \cr 0_{2n}
} \right) \eta^2  \left( \matrix{ I_{2n}   & 0_{2n}
} \right)<0.
\end{equation}
Again, using Lemma \ref{L2}, this condition is equivalent to:
\begin{equation}
 \left( \matrix{ \tilde A_n ^\dag P+P \tilde A_n  & P & I_{2n} \cr
P & -I_{2n} &0 \cr I_{2n}& 0& \frac{-1}{\eta^2} I_{2n}} \right)< 0.  
\end{equation}
This condition is equivalent to the feasibility of LMI (\ref{LMI1})-(\ref{LMI2}) along with condition $\zeta< \frac{1}{\eta^2}$, which completes the proof.
\end{proof}
Part (\ref{iiii}) of Theorem \ref{th3}, can provide an alternative viewpoint independent of norm-bounded condition (\ref{nbc}): to ensure robust stability of $\overline{G}$, the largest eigenvalues of members of $\overline{\Delta A}$ must not exceed $\zeta ^{\frac{-1}{2}}$.

\subsection{Remarks on Robust Stability Analysis}
\par In subsections \ref{Small} and \ref{Lyapunov}, two different approaches  were presented to analyze robustness of the admissible uncertain quantum networks $\overline{G}$. In both approaches, a sufficient condition for robustness of $\overline{G}$ was proposed through studying an upper-bound of the largest singular values of additive perturbations in $\overline {\Delta A}$. Moreover, equation (\ref{delta}) provides a general algebraic structure for members of $\overline {\Delta A}$ regarding different classes of uncertain QIONs. 
\par The algebraic form of additive perturbations in (\ref{delta}) includes both linear (second and third terms) and nonlinear (first term) structures with respect to uncertainty parameters ($\tilde C _\delta$ and $i \tilde{\Omega}_{\Delta H}$). Additionally, due to generality of the uncertainty decomposition scheme (Theorem \ref{th1}), proposing a unique algorithm to relate the norm-bounded condition in (\ref{nbc}) to upper bounds on uncertainty parameters is not straight-forward. In general, using the connectedness and compactness properties in Assumption \ref{ass1}, the set of additive perturbations $\overline{\Delta A}$ can be characterized by a vector of parameters $\hat \Theta$. Each member $\theta_i$ of $\hat \Theta$ can vary within an interval $[\theta_i^{min},\theta^{max}_i]$. Except for rare situations, it is straight-forward to find the bound $\eta$ in (\ref{nbc}) by examining the corner points in the set of parameters $\hat \Theta$. Comparing the bound $\eta$ with each of the three approaches outlined in Theorems \ref{th2} and \ref{th3} would conclude the robustness analysis.

 \section{Conclusion and Further Research}
In this paper, we presented two different approaches to study robust stability in linear quantum input-output networks. To this end, uncertainties in the network were decomposed into a separate sub-network. This procedure is analogous to the $P-\Delta$ decomposition in robust control theory \cite{zhou1998essentials} for linear quantum networks. The small-gain-like sufficient condition in the first approach provides the intuition for further theoretical studies on robust stability of different classes of linear QIONs. In the second method, robust stability was formulated as a minimization problem in terms of an LMI. This approach provides a straight-forward computational method for designers to analyze robust stability. 
\par One potential research path would be deriving robust control schemes for uncertain linear quantum input-output networks based on the proposed uncertainty decomposition method. Alternatively, investigating the robust stability of quantum input-output networks with non-dominant nonlinear dynamics would be another line of research.

\section{References}
\bibliographystyle{unsrt}
\bibliography{main}%

\end{document}